\documentclass[prc,notitlepage, twocolumn,
showpacs,floatfix,nofootinbib,preprintnumbers,superscriptaddress,amsmath,amssymb]{revtex4-1}

\usepackage{graphicx}\usepackage{epsfig}
\usepackage{amsmath,amssymb}
\usepackage{bm}
\usepackage{color}
\usepackage{float}
\usepackage{dcolumn} 
\usepackage{enumerate}

\begin{document}

\newcommand{\ket}[1]{\left| #1 \right\rangle}
\newcommand{\bra}[1]{\left\langle #1 \right|}
\newcommand{\Fhat}{\hat{F}}
\newcommand{\comment}[1]{{\bf #1}}

\title{Complex-energy approach to sum rules within nuclear density functional theory}
\author{Nobuo Hinohara}

\affiliation{%
Center for Computational Sciences, University of Tsukuba, Tsukuba, 305-8571, Japan
}
\affiliation{%
National Superconducting Cyclotron Laboratory, Michigan State University, East Lansing, Michigan, 48824-1321, USA
}
\affiliation{%
Joint Institute of Nuclear Physics and Applications, Oak Ridge, Tennessee 37831-6374, USA
}
\author{Markus Kortelainen}%
\affiliation{%
Department of Physics, P.O. Box 35 (YFL), FI-40014, University of Jyv\"{a}skyl\"{a}, Finland
}
\affiliation{%
Helsinki Institute of Physics, P.O. Box 64, FI-00014 University of Helsinki, Finland
}
\author{Witold Nazarewicz}
\affiliation{%
Department of Physics and Astronomy and NSCL/FRIB Laboratory, Michigan State University, East Lansing, Michigan 48824, USA
}
\affiliation{%
Physics Division, Oak Ridge National Laboratory, Oak Ridge, Tennessee 37831-6373, USA
}
\affiliation{%
Institute of Theoretical Physics, University of Warsaw, ul. Ho\.{z}a 69, PL-00-861, Warsaw, Poland
}
\author{Erik Olsen}
\affiliation{%
Department of Physics and Astronomy, University of Tennessee, Knoxville, Tennessee 37996-1200, USA
}

\date{\today}

\begin{abstract} 
\begin{description}
\item[Background]
The linear response of the nucleus to an external field contains unique information about the effective interaction, correlations governing the behavior of the many-body system, and properties of its excited states. To characterize the response, it is useful to use its energy-weighted moments, or sum rules.  By comparing computed sum rules with experimental values, the information content of the response can be 
utilized in the  optimization process of the nuclear Hamiltonian or nuclear energy density functional (EDF). But the additional information comes at a price: compared to  the ground state,
computation of excited states is  more demanding.
\item[Purpose]
To establish an efficient framework to compute energy-weighted sum rules 
of the response that is adaptable to the optimization of the nuclear EDF
and large-scale surveys of collective strength, we have developed a new technique 
within the complex-energy finite-amplitude method (FAM) based on the quasiparticle random-phase approximation (QRPA).
\item[Methods]
To compute  sum rules,
we carry out  contour integration of the  response function in the complex-energy plane. We benchmark our results against  the conventional matrix formulation of the QRPA theory, the Thouless theorem 
for the energy-weighted sum rule, and the dielectric theorem for the inverse energy-weighted sum rule.
\item[Results]
We derive the  sum-rule expressions from the contour integration of the complex-energy FAM.
We demonstrate that calculated sum-rule values agree with those obtained from  the matrix formulation of the QRPA. We also discuss the applicability of both the Thouless 
theorem about the energy-weighted sum rule and the dielectric theorem for the inverse energy-weighted sum rule to nuclear density functional theory 
in cases when the EDF is not based on a  Hamiltonian.
\item[Conclusions]
The proposed sum-rule technique based on the complex-energy FAM is a tool of choice when optimizing effective interactions or energy functionals.
The method is very efficient and well-adaptable to parallel computing. The FAM formulation is especially useful when standard theorems based on commutation relations involving the nuclear Hamiltonian and external field cannot be used.
\end{description}
\end{abstract}


\maketitle


\section{Introduction}
Atomic nuclei exhibit various kinds of collective excitations,
with characteristics considerably different from simple nucleonic excitations \cite{bohr98a,(Rin00)}.
Among those, giant resonances  form a  distinct class \cite{(Har01)}. 
Although their excitation energies are relatively high compared to the low-energy collective modes, 
the main characteristics of giant resonances are understood in terms of the superposition of many nucleonic excitations.
Experimentally, various types of giant resonances have been seen. Examples are
shape vibrations, spin excitations, and charge-exchange  excitations of various 
multipolarity and isospin.
These modes carry  rich information about basic nuclear properties.

There has been excellent progress in the modeling  of atomic nuclei  using  nuclear density functional theory (DFT) \cite{RevModPhys.75.121}.
State-of-the-art  energy density functionals (EDFs), optimized to various classes of data \cite{PhysRevC.79.034310,PhysRevC.82.024313,PhysRevC.85.024304,PhysRevC.89.054314,PhysRevC.87.044320} enable a quantitative description of global nuclear 
properties throughout the nuclear landscape \cite{(Erl12),(Afa13),PhysRevC.89.054318}.
Ground-state properties of nuclei, such as binding energies, charge radii, effective single-particle energies 
of doubly-closed shell nuclei, and basic parameters characterizing the nuclear matter equation of  state,  are typically used as empirical inputs 
in EDF parameter optimization. However, properties of excited states, such as giant resonances, are seldom considered (see Refs.~\cite{VanGiai1981379,PhysRevC.33.335,PhysRevC.60.014316,PhysRevC.72.014310, (Tri08), PhysRevC.79.034310,PhysRevC.85.035201,PhysRevC.86.031306} for representative examples of work along those lines). This results in large uncertainties of EDF parameters sensitive to, and governing, low- and high-frequency nuclear excitations. 
 The EDFs of the next generation are expected to overcome this deficiency by including  selected properties of the giant resonances into the pool of observables used in the optimization.

To extract the information content of  giant resonances, the sum rule technique 
\cite{PhysRevC.7.2281, Bertsch1975125, Lipparini1989103, Bohigas1979267,Brink1976285} has been widely used. For instance, mean giant resonance energies   can be related to the ratio of the sum rules of different energy moments \cite{Lipparini1989103,Gleissl1990205,ANDP:ANDP19925040805,PhysRevC.79.054329,PhysRevC.69.064312}.
The inverse energy-weighted sum  rule  provides information about the nuclear polarizability, which is the fundamental quantity characterizing the nuclear response. An important quantity, in the context of studies of neutron-rich matter, is the electric dipole polarizability, which  is 
related to symmetry energy 
and its density-dependence \cite{PhysRevC.81.051303,PhysRevC.85.041302,PhysRevC.87.014324}. Various polarizabilities carry information about  instabilities in  nuclear matter \cite{Stringari197687,PhysRevC.85.054317,Chamel10}.  In some cases, the Thouless theorem \cite{Thouless196178, PhysRevC.66.024309, PhysRevC.72.014304}
provides a simple way to access  sum rules directly from the Hartree-Fock-Bogoliubov (HFB) solution.
Unfortunately, the Thouless
theorem applies to positive-odd energy moments, and simple expressions can be derived only for simple operators (such as  multipole moments).
Moreover, the theorem is justified only if a Hamiltonian representation of the interaction is available, which is generally not the case for nuclear DFT where modern EDFs are usually not connected to an underlying Hamiltonian, and often break local gauge invariance \cite{Matsuo2001371}.
Therefore, an efficient technique 
to compute  nuclear sum rules, regardless of the form of the operator $\hat{F}$,  is desired. 

The direct  evaluation of  sum rules  from  self-consistent QRPA matrix solutions  is 
computationally demanding because of configuration spaces involved.
A recent formulation of the sum rule in terms of QRPA matrices enables the computation of sum rules 
without diagonalizing the  QRPA matrix \cite{PhysRevC.79.054329}. Nevertheless, this method still requires 
knowledge of the QRPA matrix, which has large dimensions, especially when spherical 
symmetry is broken. Other recent developments include applications of the Lanczos algorithm to RPA sum rules \cite{Johnson1999155} and the use of the Lorentz integral transform method and the Lanczos technique  \cite{PhysRevC.89.064317}.

The finite amplitude method \cite{nakatsukasa:024318}, based on the linear-response approach,
significantly reduces the computational cost of the QRPA problem. The residual two-body interaction is numerically 
computed from the finite-amplitude nucleonic fields induced by an external polarizing field.
The FAM has been recently implemented to  various self-consistent frameworks, including  three-dimensional HF  \cite{nakatsukasa:024318}, spherical HFB \cite{PhysRevC.84.014314}, axially-deformed Skyrme-HFB
\cite{PhysRevC.84.041305,PhysRevC.90.024308,PhysRevC.90.051304}, and relativistic mean-field models \cite{PhysRevC.87.054310,PhysRevC.88.044327}.
The FAM has been applied to the description of giant resonances and low-energy dipole strength  \cite{PhysRevC.80.044301,PhysRevC.84.021302}, the
computation of the QRPA matrix elements  \cite{PhysRevC.87.014331}, and the description of discrete low-lying QRPA 
modes by means of the  contour integration technique in the complex energy plane \cite{PhysRevC.87.064309}.

The objective  of this study is to propose an  efficient approach to sum rules  by using the contour integration technique of 
Ref.~\cite{PhysRevC.87.064309}. Because of its inherently parallel structure, the new method is ideally suited 
for   optimizations   of next-generation nuclear EDFs, 
informed by experimental data on multipole and charge-exchange strength.
This paper is organized as follows. Section~\ref{sec:basic} summarizes the basic expressions. In Sec.~\ref{sec:famsumrule}, we present the formulation of the complex-energy FAM approach to sum rules.
Section~\ref{sec:calculation} contains numerical tests, benchmarking examples, and applications to  realistic cases. The conclusions and outlook are given in Sec.~\ref{sec:conclusion}.


\section{Basic Expressions \label{sec:basic}}

\subsection{Sum rule \label{sec:sumrule}}

The ground-state (g.s.) strength function $S(E)$ for a one-body operator $\hat{F}$ is defined as
\begin{equation}\label{Sf}
S(E) \equiv \sum_\nu \delta(E-E_\nu)|\langle \nu | \hat{F} | 0 \rangle|^2,
\end{equation}
where $\ket{0}$  and  $\ket{\nu}$ denote, respectively, the ground state  and excited state of the system.
The $k$-th  moment of $S(E)$,
\begin{equation}
  m_k(\hat{F})=\int (E-E_0)^kS(E)\,dE,
\end{equation}
is called the energy-weighted sum rule of order $k$. In terms of the transition matrix elements of $\hat{F}$, it is given by:
\begin{align}
  m_k(\hat{F}) \equiv \sum_{\nu} (E_\nu - E_0)^k |\langle \nu | \hat{F} | 0 \rangle|^2.\label{eq:sumrule}
\end{align}

As discussed in, e.g., Refs.~\cite{bohr98a,(Rin00)}, certain sum rules are independent of the specific many-body theory used to describe the ground state and excited states.
For example, the nuclear shell model and QRPA frameworks have been widely used to evaluate the sum rules.
In  QRPA, the excitation energy $E_{\nu} - E_0$ is replaced with the QRPA frequency $\Omega_\nu$, 
which is the eigenvalue of the matrix equation:
\begin{align}
  \begin{pmatrix} A & B \\ -B^\ast &-A^\ast \end{pmatrix} 
  \begin{pmatrix} X^\nu \\ Y^\nu \end{pmatrix} = \Omega_\nu 
  \begin{pmatrix} X^\nu \\ Y^\nu \end{pmatrix} \, , \label{eq:QRPA}
\end{align}
where $A$ and $B$ are  QRPA matrices.
The QRPA equation (\ref{eq:QRPA}) has positive-energy solutions $\Omega_\nu>0$ $(\nu>0)$ with $(X^\nu, Y^\nu)$, and mirror negative-energy 
solutions  $\Omega_{-\nu}=-\Omega_{\nu}<0$ with $(X^{-\nu}, Y^{-\nu}) = (Y^{\nu\ast}, X^{\nu\ast})$. 
The positive frequency solutions, being the physically relevant ones, are used for the sum rule, and the summation in 
Eq.~(\ref{eq:sumrule}) is, therefore, restricted to QRPA modes with $\nu>0$.

\subsection{Finite amplitude method \label{sec:fam}}
 
The FAM is an efficient technique to obtain the response function  $S(E)$ without explicitly computing the  $A$ and $B$ QRPA matrices in Eq.~(\ref{eq:QRPA}).
For the details pertaining to FAM, we refer the reader to, e.g., Ref.~\cite{PhysRevC.84.014314}.
The complex response function for a given operator $\hat{F}$ at a given complex frequency $\omega_\gamma = \omega + i\gamma$, 
found as a solution of the FAM equations, is given as
\begin{align}
  S(\hat{F},\omega_\gamma) = 
- \sum_{\nu>0} \left\{\frac{|\langle \nu|\hat{F}|0\rangle|^2}{\Omega_\nu-\omega_\gamma}
           + \frac{|\langle 0|\hat{F}|\nu\rangle|^2}{\Omega_\nu+\omega_\gamma}\right\} \, .
\end{align}
The Lorentzian distribution of the strength function is obtained by taking the imaginary part of $S$:
\begin{eqnarray}
& &  -\frac{1}{\pi} {\rm Im}\, S(\hat{F},\omega_\gamma) \nonumber \\
& &= \frac{\gamma}{\pi} \sum_{\nu>0} \left\{\frac{|\langle \nu|\hat{F}|0\rangle|^2}{(\Omega_\nu-\omega)^2+\gamma^2}
                             - \frac{|\langle 0|\hat{F}|\nu\rangle|^2}{(\Omega_\nu+\omega)^2+\gamma^2} \right\} \, .
\end{eqnarray}
A contour integration along the path $C_{\nu}$, which encircles a real energy pole $\Omega_\nu$ of the response 
function, gives the QRPA transition strength to state $|\nu\rangle$ \cite{PhysRevC.87.064309}:
\begin{align}
 \frac{1}{2\pi i}  \oint_{C_\nu} S(\hat{F},\omega_\gamma) d\omega_\gamma = |\langle \nu|\hat{F}|0\rangle|^2 \quad  (\Omega_\nu>0) \, ,
\end{align}
or, alternatively, along $C_{-\nu}$, 
\begin{align}
  \frac{1}{2\pi i}  \oint_{C_{-\nu}} S(\hat{F},\omega_\gamma) d\omega_\gamma =& - |\langle 0|\hat{F}|\nu\rangle|^2  \nonumber \\
  =& -|\langle \nu|\hat{F}^\dagger|0\rangle|^2 \quad (\Omega_{-\nu}<0).
\end{align}
For a small $\gamma\ll \omega$, the relation $1/(\omega+i\gamma)=P(1/\omega) - i\pi \delta(\omega)$ holds,
and the sum rules can be formally calculated using
\begin{align} \label{eq:mkgamma}
  m_k(\hat{F}) = 
  -\frac{1}{\pi} \lim_{\gamma\to 0}\int_0^\infty \omega^k {\rm Im}\, & S(\hat{F},\omega+i\gamma) d\omega \, .
\end{align}
An approximate value of the sum rules can be found from this expression from a finite value of $\gamma$ \cite{PhysRevC.84.014314,PhysRevC.88.044327,PhysRevC.84.021302,PhysRevC.84.041305}.
However, to guarantee sufficient
numerical accuracy, a very fine mesh would be required for the integration (\ref{eq:mkgamma}) to take into account all the
QRPA modes, whose locations are not known beforehand.


\section{Sum rule expressions in FAM \label{sec:famsumrule}}

In this section we introduce the sum rule approach based on the contour integration of the FAM.
For simplicity, we assume that the operator $\hat{F}$ cannot excite spurious modes, and all the 
QRPA energies $\Omega_\nu$ are non-zero.
We also assume that the HFB state is stable with respect to  small density variations, i.e.,
there are no imaginary-frequency QRPA solutions. This guarantees that all the QRPA poles 
$\Omega_\nu$ lie on the real axis. In the following,  we shall adopt the notation $\omega$ for a complex frequency.

The basic idea behind the FAM approach to  sum rules is to utilize the identity based on Cauchy's integral theorem: 
\begin{align}
  \oint_D f(\omega) S(\hat{F},\omega) d\omega = \sum_{\nu>0} f(\Omega_\nu) |\langle\nu|\hat{F}|0\rangle|^2,\label{eq:D}
\end{align}
where the contour $D$ encircles all the positive QRPA frequencies $\Omega_\nu>0$,
and excludes all the singularities of the complex function $f(\omega)$.
By setting $f(\omega)=\omega^k$, we obtain the expressions for the sum rule $m_k(\hat{F})$.

In the following, we assume the operator $\hat{F}$ to be Hermitian for simplicity.
In this case, positive and negative energy solutions are associated with the same transition strength:
\begin{align}
  |\langle \nu| \hat{F}|0\rangle|^2 =   |\langle 0| \hat{F}|\nu\rangle|^2 \,.
\end{align} 
The above equation does not hold when $\hat{F}$ is not  Hermitian. However,  Eq.~(\ref{eq:D}) still can be used with
an appropriately chosen contour $D$.

\subsection{Laurent series of the FAM response function}

By using the Laurent series expansion of $(1-z)^{-1}$, 
we can derive the expansion of the FAM response function.
The FAM response function  has poles at $\omega = \Omega_\nu$ and $-\Omega_\nu$.
In the inner region below the lowest QRPA pole,  $|\omega|<\min_{\nu>0}\Omega_\nu$, $S(\hat{F},\omega)$ can be written as 
\begin{align}
  S(\hat{F},\omega) = -2 \sum_{n=0}^\infty m_{-(2n+1)}(\hat{F})\omega^{2n}.  \label{eq:Laurent_in}
\end{align}
One can see that odd-$k$ sum rules can be simply related to the expansion coefficients  of (\ref{eq:Laurent_in}). The same is true 
in the outer region above the highest QRPA pole, $|\omega|> \max_{\nu>0} \Omega_\nu$, where the response function can be expanded as
\begin{align}
  S(\hat{F},\omega) = 2 \sum_{n=0}^\infty \frac{m_{2n+1}(\hat{F})}{\omega^{2n+2}}. \label{eq:Laurent_out}
\end{align}

\begin{figure}[htb]
\includegraphics[width=0.8\columnwidth]{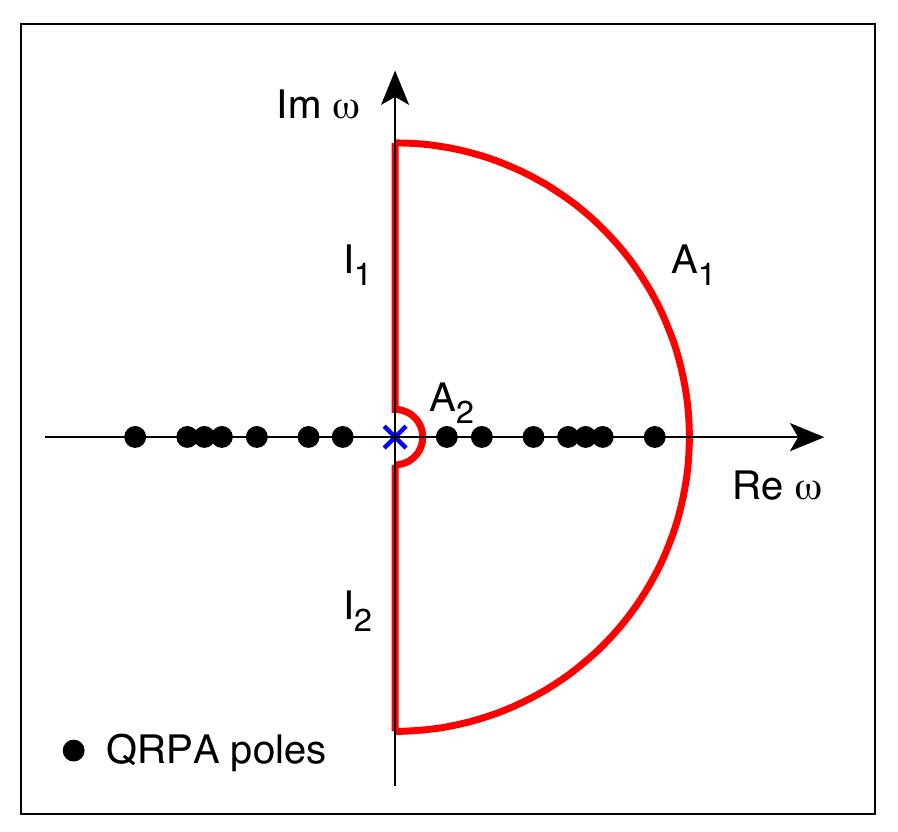}
\caption{\label{fig:contourD} (Color online)
The contour $D$ (oriented counterclockwise) in the complex-$\omega$ plane  used to evaluate sum rules. The contour consists of two semicircles $A_1$ and $A_2$ 
(of radii $R_{A_1}$ and $R_{A_2}$, respectively) and two segments  $I_1$ and $I_2$ on the imaginary axis. The positive 
QRPA  poles are all located between $R_{A_2}$ and $R_{A_1}$.}
\end{figure}

The expansions (\ref{eq:Laurent_in}) and (\ref{eq:Laurent_out}) are generalizations of expansions proposed in Ref.~\cite{Bohigas1979267} to the full complex energy plane.
We note that the inverse energy-weighted sum rule ($k=-1$) is found by setting $\omega=0$ in Eq.~(\ref{eq:Laurent_in}). 
This should be done with care, however.  If spurious modes are present, they would produce a zero-frequency pole resulting in numerical instabilities near or at the pole.
If we choose the semicircle $A_1$ (counterclockwise) and $A_2$ (clockwise) with the radii
satisfying $0 < R_{A_2} < \min_{\nu>0}\Omega_\nu$ and $R_{A_1} > \max_{\nu>0}\Omega_\nu$, as in Fig.~\ref{fig:contourD}, 
we can apply the series (\ref{eq:Laurent_in}) and (\ref{eq:Laurent_out}) along the integration path.
The odd-$k$ sum rules are then given as:
\begin{align}
  m_{k}(\hat{F}) =\left\{ \begin{array}{ll}
  \displaystyle\frac{1}{2\pi i}\int_{A_1} \omega^k S(\hat{F},\omega) d\omega & (k>0, {\rm odd}), \\
  \displaystyle\frac{1}{2\pi i}\int_{A_2} \omega^k S(\hat{F},\omega) d\omega & (k<0, {\rm odd}). \end{array}\right.
\end{align}
To evaluate even-$k$ sum rules, we need to connect $A_1$ and $A_2$ to enclose the positive-energy poles. To this end, we consider  contour $D$ of Fig.~\ref{fig:contourD} composed of semicircles $A_1$, $A_2$ connected by straight segments $I_1$ and $I_2$  on the imaginary axis.

In summary, regardless of the moment $k$,  the sum rule is given by the integration along  $D$:
\begin{align}
m_k(\hat{F}) = \frac{1}{2\pi i}  \oint_D \omega^k S(\hat{F},\omega) d\omega = \sum_{\nu>0} \Omega_\nu^k |\langle \nu|\hat{F}|0\rangle|^2. \label{eq:sumruleforallk}
\end{align}
However, for certain moments $k$, some parts of path $D$ do not contribute to the sum rule. 
For odd values of $k$,  the contributions from $I_1$ and $I_2$  cancel each other. Furthermore, for negative 
$k$, application of Jordan's lemma, together with a limit of $R_{A_1}\to\infty$, allows for the removal of the contribution from $A_1$. 
For positive $k$, there is no pole at $\omega=0$, and the limit $R_{A_2}\to 0$ can be taken.
Table~\ref{table:contour} lists the portions of the contour $D$ required for each $k$. Furthermore, 
for even $k$, the contributions from $I_1$ and $I_2$  are identical.
Similar contours were considered  in Ref.~\cite{TMF32_134} to compute energy-weighted sum rules.

\begin{table}[h]
\caption{Portions of the contour $D$ required for computing various sum rules $m_k$. For sum rules with even $k$, the contributions from  $I_1$ and $I_2$  are identical.
\label{table:contour}}
\begin{ruledtabular}
\begin{tabular}{lr}
$k$ & Required portions of $D$\\
\hline
negative, even &  $A_2$, $I_1$, $I_2$ ($R_{A_1}\to\infty$) \\
negative, odd &   $A_2$  ($R_{A_1}\to\infty$) \\
0             &   $A_1$, $A_2$, $I_1$, $I_2$ \\
positive, odd &   $A_1$ ($R_{A_2}\to 0$) \\
positive, even &  $A_1$, $I_1$, $I_2$ ($R_{A_2}\to 0$) 
\end{tabular}
\end{ruledtabular}
\end{table}


\section{Results \label{sec:calculation}}

\subsection{Numerical checks and benchmarking against MQRPA}

To check the FAM  approach to sum rules, following Refs.~\cite{PhysRevC.84.041305,PhysRevC.87.064309}
we consider the oblate configuration of $^{24}$Mg computed with  the SLy4~\cite{SLy4} Skyrme EDF.
The HFB calculations were carried out with the DFT solver HFBTHO \cite{Stoitsov20131592} in a model space of five 
harmonic oscillator  shells by employing a volume pairing with the strength of $V_0=-125.20\,{\rm MeV\,fm}^{3}$ 
and a 60\,MeV quasiparticle energy cutoff.
The resulting  oblate minimum of $^{24}$Mg  has nonzero pairing in protons and neutrons.
The small single-particle model space employed makes it possible to benchmark FAM results against  the matrix formulation of QRPA 
(MQRPA) \cite{PhysRevC.81.064307} without any further truncation.
To compute spatial integrals we used 
Gauss-Hermite ($N_{\rm GH}=30$), Gauss-Laguerre ($N_{\rm GL}=30$), 
and Gauss-Legendre ($N_{\rm Leg}=30$) quadratures.
The finite-amplitude expansion parameter $\eta$ was set to be $10^{-7}$, and the convergence criterion of the FAM was set
such that
the change of the individual FAM amplitudes from the previous iteration should be less than $10^{-5}$.
The integration along semicircles $A_1$ and $A_2$ was discretized with $N_{A_1}$ and $N_{A_2}$ points, respectively.
In addition, the integration along $I_1$ was discretized with $N_{I_1}$ points and evaluated using the composite Simpson's rule.
As for  negative-$k$ moments, the composite Simpson's rule was applied to the variable $1/\omega$ to describe the divergent 
behavior of integrand around $\omega=0$. 
In  this particular test case, the smallest and largest energy MQRPA poles appear at 1.3\,MeV and 128.7\,MeV, respectively. Consequently,
the contour radii were set to $R_{A_2} = 1$\,MeV and $R_{A_1}=200$\,MeV.
In order to systematically assess our numerical procedure for different moments $k$, we used the same contour $D$ for 
all cases, without simplifications listed in Table~\ref{table:contour}.

 As far as  the external field $\hat{F}$ is concerned,  we considered the  isoscalar (IS) and isovector (IV)  monopole (M) and quadrupole (Q) operators:
\begin{align}
  \hat{F}^{\rm ISM} =& \frac{eZ}{A} \sum_{i=1}^A r^2_i \, , \label{eq:ISM}\\
  \hat{F}^{\rm IVM} =& \sum_{i=1}^A e_{\rm eff}(\tau_{3i})r^2_i\hat{\tau}_{3i},\\
  \hat{F}^{\rm ISQ} =& \frac{eZ}{A} \sum_{i=1}^A r^2_i Y_{20}(\theta_i,\varphi_i) \, , \\
  \hat{F}^{\rm IVQ} =& \sum_{i=1}^A e_{\rm eff}(\tau_{3i}) r^2_i Y_{20}(\theta_i,\varphi_i)\hat{\tau}_{3i}, \label{eq:IVQ}
\end{align}
where
$e_{\rm eff}({\rm n})=eZ/A$ and $e_{\rm eff}({\rm p})=-eN/A$.

It is worth noting that neutron and proton 
pairing-rotational spurious modes, associated with the breaking of the particle number symmetry, are present in  the  $K^\pi=0^+$ sector. Fortunately,
these modes -- generated by the neutron and proton particle number operators -- cannot be excited by 
 particle-hole  operators (\ref{eq:ISM})-(\ref{eq:IVQ}). Therefore,  the presence of  pairing-rotational spurious modes does not cause any additional difficulties \cite{nakatsukasa:024318,PhysRevC.84.041305}.
 \begin{table*}[h]
\caption{The real part of the integral (\ref{eq:sumruleforallk}) 
 for $-4\le k\le 4$  and  $\hat{F}^{\rm ISM}$ (in
${\rm MeV}^k\,{\rm e}^2\,{\rm fm}^4$)
along 
semicircle $A_1$  with $R_{A_1}=200$\,MeV. The integral was discretized with  $N_{A_1}$ points.
The numbers in parentheses denote powers of 10.
\label{table:A1}}
\begin{ruledtabular}
\begin{tabular}{cccccccccc}
$N_{A_1}$ & $k=-4$ & $k=-3$ & $k=-2$ & $k=-1$ & $k=0$ & $k=1$ & $k=2$ & $k=3$ & $k=4$ \\ \hline
  2 & -9.0(-9) & -2.6(-6)  & -3.8(-4) & -2.7(-3) & 14.4510 & 4195.31 & 608575 & 4318446 & -23121542422 \\
  3 &  9.0(-9) &  6.8(-8)  & -1.7(-4) &  1.1(-4) & 13.8328 & 4199.78 & 574147 & 4338677 & -10598058261 \\
  4 &  3.3(-9) & -2.6(-9)  & -1.4(-4) & -5.7(-6) & 13.5751 & 4199.39 & 562629 & 4336463 &  -8502934038 \\
  5 &  2.5(-9) &  2.2(-10) & -1.3(-4) &  4.1(-6) & 13.4599 & 4199.44 & 557421 & 4336545 &  -7722808528 \\
  7 &  2.0(-9) &  7.1(-12) & -1.2(-4) &  2.9(-6) & 13.3593 & 4199.44 & 552937 & 4336634 &  -7119788306 \\
  9 &  1.9(-9) &  2.1(-12) & -1.1(-4) &  2.9(-6) & 13.3181 & 4199.44 & 551106 & 4336643 &  -6889795483 \\
 10 &  1.8(-9) &  1.9(-12) & -1.1(-4) &  2.9(-6) & 13.3061 & 4199.44 & 550575 & 4336644 &  -6824728295 \\
 11 &  1.8(-9) &  1.9(-12) & -1.1(-4) &  2.9(-6) & 13.2972 & 4199.44 & 550183 & 4336638 &  -6777106568 \\
 12 &  1.8(-9) &  2.0(-12) & -1.1(-4) &  2.9(-6) & 13.2905 & 4199.44 & 549885 & 4336648 &  -6741178415 \\
101 &  1.6(-9) &  1.9(-12) & -1.1(-4) &  2.9(-6) & 13.2556 & 4199.44 & 548341 & 4336643 &  -6558793102 \\
201 &  1.6(-9) &  1.9(-12) & -1.1(-4) &  2.9(-6) & 13.2552 & 4199.44 & 548325 & 4336643 &  -6556876296 \\
301 &  1.6(-9) &  2.0(-12) & -1.1(-4) &  2.9(-6) & 13.2551 & 4199.44 & 548322 & 4336644 &  -6556517206 
\end{tabular}
\end{ruledtabular}
\end{table*}

To begin with, we checked the convergence of the integral  (\ref{eq:sumruleforallk})  along $A_1$, $A_2$, and $I_1$ with respect of the 
number of integration points. The results are presented in Tables~\ref{table:A1}-\ref{table:I1} for the isoscalar monopole operator.
As seen in  Table~\ref{table:A1}, the integrals along $A_1$ are small for negative $k$.
Analytically, these values should be  zero for negative-odd values of $k$; hence,
nonzero  values in  Table~\ref{table:A1} reflect the numerical error of calculations.
As far as the positive $k$ moments are concerned, the convergence is faster for odd-$k$ sum rules. In particular, 
the convergence for $k=1$ is excellent, since a 6-digit accuracy is achieved already with $N_{A_1}=5$.
The integration along $A_1$ captures the total $m_1$ and $m_3$  sum rules;
the result in Table~\ref{table:A1} indicates  that these sum rules can be computed very efficiently.
Moreover, since the semicircle $A_1$  is located very far from the QRPA poles, FAM calculations along $A_1$   converge very quickly, typically after 6 iterations. Furthermore, each FAM calculation at a given $\omega$ along the contour is  easily parallelizable;
this could significantly reduce the total computational time, although not so many points are required for the 
convergence of  $m_1$ and $m_3$.

\begin{table*}[htb]
\caption{Similar to Table~\ref{table:A1} but for the integration along $A_2$ for several values 
of $N_{A_2}$ with $R_{A_2}=1.0\,{\rm MeV}$.
\label{table:A2}}
\begin{ruledtabular}
\begin{tabular}{cccccccccc}
$N_{A_2}$ & $k=-4$ & $k=-3$ & $k=-2$ & $k=-1$ & $k=0$ & $k=1$ & $k=2$ & $k=3$ & $k=4$ \\ \hline
  5 & -1.20929 & 0.0372710 & 3.26165 & 5.00438 & 3.23108 & 1.0(-3) & -1.22915 &  2.2(-3) & 0.996926 \\
 11 & -1.06784 & 0.0369978 & 3.21892 & 5.00387 & 3.18902 & 2.7(-5) & -1.09039 &  4.1(-5) & 0.691152 \\   
 15 & -1.05236 & 0.0369925 & 3.21386 & 5.00386 & 3.18407 & 7.1(-6) & -1.07507 &  3.9(-6) & 0.663895 \\
 16 & -1.05021 & 0.0369910 & 3.21315 & 5.00385 & 3.18338 & 4.1(-6) & -1.07293 & -7.8(-7) & 0.660178 \\
 21 & -1.04368 & 0.0369929 & 3.21099 & 5.00385 & 3.18127 & 6.2(-6) & -1.06646 & -6.4(-7) & 0.649056 \\
 31 & -1.03883 & 0.0369918 & 3.20937 & 5.00385 & 3.17968 & 6.1(-6) & -1.06164 &  9.9(-7) & 0.640898 \\
 51 & -1.03625 & 0.0369916 & 3.20851 & 5.00385 & 3.17884 & 5.6(-6) & -1.05908 &  1.0(-6) & 0.636594 \\
101 & -1.03512 & 0.0369918 & 3.20813 & 5.00385 & 3.17847 & 5.9(-6) & -1.05797 &  7.2(-7) & 0.634728
\end{tabular}
\end{ruledtabular}
\end{table*}

\begin{table*}[htb]
\caption{Similar to Table~\ref{table:A1} but for the integration along $I_1$ for several values 
of $N_{I_1}$ with $R_{A_1}=200\,{\rm MeV}$ and $R_{A_2}=1.0\,{\rm MeV}$.
\label{table:I1}}
\begin{ruledtabular}
\begin{tabular}{cccccccccc}
$N_{I_1}$ & $k=-4$ & $k=-3$ & $k=-2$ & $k=-1$ & $k=0$ & $k=1$ & $k=2$ & $k=3$ & $k=4$ \\ \hline
10  & 0.523854 & -6.6(-8) & -1.472654 &  1.3(-7) & 61.4767 & -2.0(-6) & -208884 &  2.3(-2) & 3356473886 \\
30  & 0.523849 &  1.2(-6) & -1.478626 & -1.6(-6) & 61.7093 & -6.0(-7) & -208523 & -4.2(-3) & 3356786248 \\
50  & 0.523850 & -1.1(-6) & -1.477671 &  1.4(-6) & 61.7024 & -2.2(-6) & -208522 &  3.3(-2) & 3356787504 \\
100 & 0.523850 & -1.5(-6) & -1.477427 &  2.0(-6) & 61.7044 & -1.4(-6) & -208522 &  2.8(-2) & 3356787251 \\
200 & 0.523850 & -8.7(-7) & -1.477417 &  1.2(-6) & 61.7052 & -1.8(-6) & -208522 &  2.2(-2) & 3356787739
\end{tabular}
\end{ruledtabular}
\end{table*}
Table~\ref{table:A2} shows the convergence of the integral (\ref{eq:sumruleforallk})  along $A_2$.
This portion of the contour is required for the sum rules with negative $k$. Of most practical importance  is the inverse energy-weighted 
sum rule $m_{-1}$. The value of $m_{-1}$ converges here with $N_{A_2}=16$ points. In general, as compared to integration along $A_1$, more FAM iterations are required to achieve reasonable 
convergence along $A_2$. In the  case considered, typically 50 FAM iterations are  necessary for each $\omega$.
When choosing  $R_{A_2}$ one has to keep in mind
that its value
should be smaller than the lowest QRPA pole, whose energy is not a priori known. At the same time,  the convergence of  FAM calculations for negative-$k$ moments deteriorates rapidly  when $R_{A_2}$ gets too close to zero.

Table~\ref{table:I1} illustrates  the convergence along the segment $I_1$ on the imaginary axis.
As  discussed, this integration should be nonzero only for even-$k$ moments. 
The convergence for $k=0$ is reached rather  slowly, especially when  compared with the $k=4$ and $-4$ cases. This is because  the
Simpson's formula used approximates the integrand with quadratic functions, which is a poor ansatz for $k=0$.

\begin{table*}[h]
\caption{Sum rules  (in ${\rm MeV}^k\,{\rm e}^2\,{\rm fm}^4$) for the isoscalar and isovector monopole operators calculated with
MQRPA and FAM. 
The FAM calculations were performed by using $N_{A_1}=301$, $N_{A_2}=101$, and $N_{I_1}=200$ integration points. 
\label{table:ismsr}}
\begin{ruledtabular}
\begin{tabular}{cccccccccc}
  & $k=-4$ & $k=-3$ & $k=-2$ & $k=-1$ & $k=0$ & $k=1$ & $k=2$ & $k=3$ & $k=4$ \\ \hline \\[-6pt]
MQRPA(ISM) & 0.013077  & 0.037185  & 0.253118 & 5.00072 & 139.825 & 4200.82 & 131368 & 4342358 & 157906069 \\ 
FAM(ISM)   & 0.012579 & 0.036992 & 0.253186 & 5.00385 & 139.844 & 4199.44 & 131277 & 4336644 & 157058272 \\[4pt]
MQRPA(IVM) & 0.00063616 & 0.00273872 & 0.07120949 & 2.78540 & 113.908 & 4735.03 & 199525 & 8527358 & 370625216 \\ 
FAM(IVM)   & 0.00043157 & 0.00275227 & 0.07133510 & 2.78615 & 113.908 & 4734.30 & 199510 & 8524830 & 368643941 
\end{tabular}
\end{ruledtabular}
\end{table*}

To benchmark our FAM approach, in Table~\ref{table:ismsr} we display
the values of sum rules for the isoscalar and isovector monopole operators; they are compared with the MQRPA results based on the direct evaluation of the r.h.s. of Eq.~(\ref{eq:sumruleforallk}). Overall, there is an excellent agreement between the two sets of calculations.  This result indicates that the proposed FAM  technique can be used to predict  sum rules  of interest  in model spaces that are  too large to be treated with  
MQRPA. The convergence of integration along $A_2$ is not sufficient in the case of $k=-4$; this sum rule is, however,  less important  than  other  moments discussed.

\subsection{Thouless theorem for energy-weighted sum rule}

The Thouless theorem  \cite{Thouless196178} gives the relation between the energy-weighted sum rule $m_1(\hat{F})$
for isoscalar  $\hat{F}=\sum_{i=1}^A f(\hat{\bm{r}}_i)$ or
isovector  $\hat{F}=\sum_{i=1}^A f(\hat{\bm{r}}_i)\hat{\tau}_{3i}$ one-body operators
and the expectation value of the double commutator at the ground state   \cite{Bertsch1975125,Lipparini1989103,Bohigas1979267,Yan87}:
\begin{align}
  m_1(\hat{F}) =& \frac{1}{2}\langle [\hat{F}, [\hat{H}, \hat{F}]] \rangle
  = \frac{1}{2}(1+\kappa) \langle [\hat{F}, [\hat{T}, \hat{F}]] \rangle \nonumber \\
  =&  \displaystyle (1+\kappa) \frac{\hbar^2}{2m} \int |\nabla f(\bm{r})|^2 \rho(\bm{r})d\bm{r}
  \, , \label{eq:EWSR}
\end{align}
where $\hat{T}$ is the kinetic energy operator  and $\kappa$ is the enhancement factor, which is present in the case when $\hat{F}$ is an isovector operator.
The explicit expressions for the r.h.s. of Eq.~(\ref{eq:EWSR}) for the operators (\ref{eq:ISM})-(\ref{eq:IVQ}) are given in Appendix~\ref{sec:appendix}.

The theorem is exact when both the ground state and the excited states are many-body shell-model states, and 
has been proven for  HF+RPA  \cite{Thouless196178},  HFB+QRPA  \cite{PhysRevC.66.024309}, and second RPA \cite{PhysRevC.90.024305}.
In the case of HF+RPA, expression (\ref{eq:EWSR}) also holds for the Skyrme force due to the $\delta$-character 
of the momentum-dependent terms \cite{Brink1976285,Bohigas1979267}.
However, as pointed out in Ref.~\cite{PhysRevC.85.054317}, the theorem has not been proven for a generalized
EDF, which is not explicitly related to an effective interaction.
Deviations from relation  (\ref{eq:EWSR}) 
can be caused by, e.g.,  different assumptions about particle-hole and pairing channels, the Slater approximation to the  Coulomb exchange term, approximations to spin-orbit and tensor terms  \cite{PhysRevC.90.054303}, and generalized density dependence \cite{Rai11,Dob12,Sadoudi13}.
To the best of our knowledge, the Thouless theorem has not been proven in the case of generalized EDFs.

\begin{table*}[htb]
  \caption{The energy weighted $K^\pi=0^+$  sum rule (in  ${\rm MeV\,e}^2\,{\rm fm}^4$)
 for the operators (\ref{eq:ISM})-(\ref{eq:IVQ}) at
     the oblate minimum of $^{24}$Mg as a function of
$N_{\rm sh}$.
    The FAM values were obtained by taking $R_{A_1}=200$\,MeV and $N_{A_1}=12$; they are compared to HFB values (\ref{eq:EWSR}).
    The results without time-odd terms except for the current-current coupling 
    ($C_t^j\ne 0$ and $C_t^s(\rho_0)=C_t^{\Delta s}=C_t^{\nabla j}=C^T_t=0$) $(a)$, and
    with the full time-odd functional except for the current-current and kinetic spin-spin couplings 
    ($C_t^j=C^T_t= 0, C_t^s(\rho_0)\ne 0, C_t^{\Delta s}\ne 0$, and $C_t^{\nabla j}\ne 0$) $(b)$,
    obtained with $N_{\rm sh}$=20, are also listed.
    \label{table:ewsr}}
  \begin{ruledtabular}
    \begin{tabular}{ccccccccc} $N_{\rm sh}$
        & FAM(ISM) & HFB(ISM) & FAM(IVM) & HFB(IVM) & FAM(ISQ) & HFB(ISQ) & FAM(IVQ) & HFB(IVQ)\\ \hline
      5 & 4199.44  & 4303.67 & 4734.25 & 4752.37 & 762.638 & 767.933 & 848.110 & 845.235 \\ 
     10 & 4524.39  & 4502.75 & 4970.34 & 4940.08 & 779.019 & 775.724 & 852.995 & 849.015 \\ 
     15 & 4521.39  & 4523.80 & 4958.02 & 4960.52 & 776.587 & 776.161 & 849.482 & 849.116 \\ 
     20 & 4530.01  & 4529.46 & 4966.98 & 4966.01 & 777.425 & 776.832 & 850.145 & 849.747 \\ \hline 
     20$^{(a)}$ & 4530.07 & - & 4966.98 & -       & 777.506 & -       & 850.132 & -\\
     20$^{(b)}$ & 5297.64 & - & 5298.46 & -       & 905.461 & -       & 905.441& -
    \end{tabular} 
    \end{ruledtabular}
\end{table*}
In the following, we refer to  the value (\ref{eq:EWSR}) as the ``HFB value'' of the 
energy-weighted sum rule. In Table~\ref{table:ewsr} the energy-weighted sum rules obtained in  HFB and FAM 
are compared for different  model spaces given by $N_{\rm sh}$. In a small model space of $N_{\rm sh}=5$,  the difference between FAM and HFB values is non-negligible but
quickly becomes small with  $N_{\rm sh}$. This can be attributed to a poor representation of the operator $\Fhat$ in  small basis spaces, resulting in an error on 
the derivative of the function $f(\hat{\bm{r}})$ in Eq.~(\ref{eq:EWSR}).
In spite of the fact that the SLy4 EDF combined with volume pairing cannot be related to a force,
the numerical test in Table~\ref{table:ewsr}  demonstrates that the Thouless theorem provides a reasonably good approximation to the value of the sum rule $m_1$ for 
the Skyrme EDF.

In the notation of Ref. \cite{PhysRevC.52.1827}, the  time-odd part of the Skyrme EDF reads:
\begin{align} \label{eq:skyrmeodd}
  {\cal E}_{\rm odd} = & \sum_{t=0,1} \big[
  C^s_t(\rho_0) \bm{s}^2_t + C^{\Delta s}_t \bm{s}_t \cdot \Delta \bm{s}_t  \nonumber \\
  & +C^j_t \bm{j}^2_t + C^{\nabla j}_t \bm{s}_t \cdot (\nabla\times\bm{j}_t) +C^{T}_{t}\bm{s}_t \cdot \bm{T}_{t} \big] \, .
\end{align}
By taking the Skyrme interaction as a starting point, the time-odd and time-even coupling constants of the Skyrme EDF are 
related to each other. That is, by fixing time-even coupling constants, the time-odd part becomes also determined.
This choice also guarantees the EDF's gauge invariance \cite{(Per04)}. In the EDF picture, however, the time-odd coupling
constants could be treated as independent parameters, where some of them can be constrained by
local gauge invariance \cite{PhysRevC.65.054322,PhysRevC.52.1827}. 
With local gauge invariance assumed and tensor terms excluded, the last term of Eq.~(\ref{eq:skyrmeodd}), proportional to $C^{T}_{t}$, 
vanishes. In standard  HFB calculations for  even-even nuclei, the time-odd fields do not 
contribute because of time-reversal symmetry; hence, the time-odd part (\ref{eq:skyrmeodd})  does not affect the HFB value~(\ref{eq:EWSR}).
However, when time-reversal symmetry becomes broken, as in the case of FAM calculations, time-odd terms become active.

As shown in Table~\ref{table:ewsr}, the inclusion of the current-current term $C^j_t\bm{j}_{t}^2$ is necessary in the FAM to recover  
the HFB value of the energy-weighted sum rule of the monopole and quadrupole operators. This indicates that  the gauge invariance of the term $\rho\tau-\bm{j}^2$ should not be broken when applying the Thouless theorem to QRPA sum rules.
Other terms in the time-odd functional do not impact the energy-weighed sum rule.
Local gauge invariance also couples the  $C_t^{\nabla j}$ and  $C_t^{\nabla J}$ terms,
but the numerical results demonstrate  that these do not affect the energy-weighted sum rule.

\subsection{Dielectric theorem for the inverse energy-weighted sum rule}

The dielectric theorem connects the inverse-energy-weighted sum rule (related to nuclear polarizability) 
with the constrained potential energy surface. This theorem was proposed in Refs.~\cite{PhysRevC.7.2281,Bohigas1979267} for the HF case,
and has been proven in the HFB framework in Ref.~\cite{PhysRevC.79.054329}.
Based on this theorem, the inverse energy-weighted sum rule $m_{-1}$ can be obtained  from the curvature of the total energy ${\cal E}$ at equilibrium:
\begin{align}\label{diel}
  m_{-1}(\hat{F}) =  & \frac{1}{2} \left. \frac{\partial^2}{\partial\lambda^2} {\cal E}(\lambda)\right|_{\lambda=0} \nonumber \\
=  & \left. \frac{1}{2} \frac{\partial\bra{\phi(\lambda)}\Fhat\ket{\phi(\lambda)}}{\partial\lambda}\right|_{\lambda=0},
\end{align}
where the constrained HFB state $\ket{\phi(\lambda)}$  is
obtained by minimizing the total Routhian  containing  a linear constraint $-\lambda \hat{F}$.
We use the  relation (\ref{diel})  to compute the $m_{-1}$ sum rule.
The derivative is evaluated with a finite difference of $\Delta\lambda\,=\,0.0001\,{\rm MeV\,e}^{-1}\,{\rm fm}^{-2}$.
The resulting $m_{-1}$ values are  compared with those from the FAM in
Table~\ref{table:mm1}. A good agreement is  found already in a small   model space ($N_{\rm sh}=5$) where $m_{-1}$ is not fully converged, indicating
that the dielectric theorem works well, independently of the size of the model space.
This finding is consistent with the proof of Ref.~\cite{PhysRevC.79.054329}, which applies to an arbitrary  size of quasiparticle  space.

\begin{table*}
  \caption{Inverse energy-weighted sum rule (in  ${\rm MeV}^{-1}\,{\rm e}^2\,{\rm fm}^4$)
    computed using the dielectric theorem (HFB) and the FAM for various sizes of the model space given by $N_{\rm sh}$. FAM calculations were performed using 
    $N_{A_2}=22$ and  $R_{A_2}=1.0$\,MeV.\label{table:mm1}}
  \begin{ruledtabular}
    \begin{tabular}{ccccccccc}
$N_{\rm sh}$ & FAM(ISM) & HFB(ISM) & FAM(IVM) & HFB(IVM) & FAM(ISQ) & HFB(ISQ) & FAM(IVQ) & HFB(IVQ)\\ \hline
      5  & 5.00385 & 5.00375 & 2.78615 & 2.78614 & 4.44830 & 4.44765 & 0.798680 & 0.798680 \\
      10 & 11.2033 & 11.2102 & 5.09467 & 5.09671 & 5.21547 & 5.21586 & 1.07516 & 1.07524 \\
      15 & 12.4930 & 12.5009 & 5.71677 & 5.71960 & 5.31250 & 5.31268 & 1.12916 & 1.12910 \\
      20 & 12.9506 & 12.9634 & 6.06842 & 6.07304 & 5.35499 & 5.35730 & 1.15744 & 1.15771
    \end{tabular}
    \end{ruledtabular}
\end{table*}


\subsection{Example of systematic calculations \label{sec:systematic}}

\begin{table*}
  \caption{Isoscalar monopole and quadrupole energy-weighted $K^\pi=0^+$ sum rules in units of ${\rm MeV\,e}^2\,{\rm fm}^4$ 
   computed with the FAM and the HFB techniques for $^{142-152}$Nd and $^{144-154}$Sm isotopes. 
   The quadrupole deformation $\beta$, neutron and proton pairing gaps ($\Delta_{\rm n}$ and
   $\Delta_{\rm p}$, respectively), and total rms radius $\sqrt{\langle r^2\rangle}$ are also shown. \label{table:NdSm}}
  \begin{ruledtabular}
  \begin{tabular}{ccccccccc}
    & $\beta$ & $\Delta_{\rm n}$(MeV) & $\Delta_{\rm p}$(MeV) & $\sqrt{\langle r^2\rangle}$(fm) & HFB(ISM) & FAM(ISM) & HFB(ISQ) & FAM(ISQ) \\ \hline\\[-6pt]
    $^{142}$Nd & 0.0     & 0.00 & 1.21 & 4.92 & 50497 & 50724 & 10046 & 10068 \\
    $^{144}$Nd & 0.09 & 0.49 & 1.09 & 4.95 & 50453 & 50647 & 10606 & 10626 \\
    $^{146}$Nd & 0.15 & 0.55 & 1.00 & 4.99 & 50590 & 50769 & 11042 & 11062 \\
    $^{148}$Nd & 0.21 & 0.00 & 0.93 & 5.03 & 50788 & 50936 & 11412 & 11429 \\
    $^{150}$Nd & 0.31 & 0.64 & 0.00 & 5.11 & 51667 & 51806 & 12287 & 12301 \\
    $^{152}$Nd & 0.32 & 0.00 & 0.00 & 5.14 & 51649 & 51762 & 12375 & 12383 \\ \\[-6pt]
    $^{144}$Sm & 0.0      & 0.00 & 1.10 & 4.94 & 53635 & 53873 & 10670 & 10693 \\
    $^{146}$Sm & 0.06 & 0.55 & 1.08 & 4.97 & 53492 & 53712 & 11048 & 11069 \\
    $^{148}$Sm & 0.16 & 0.56 & 1.07 & 5.01 & 53754 & 53957 & 11770 & 11792 \\
    $^{150}$Sm & 0.21 & 0.16 & 0.93 & 5.06 & 53979 & 54145 & 12189 & 12207 \\
    $^{152}$Sm & 0.28 & 0.57 & 0.69 & 5.11 & 54474 & 54646 & 12768 & 12786 \\
    $^{154}$Sm & 0.32 & 0.09 & 0.65 & 5.16 & 54707 & 54849 & 13071 & 13084
  \end{tabular}
  \end{ruledtabular}
\end{table*}
As an illustrative example, we discuss the energy-weighted  $K^\pi=0^+$  sum rules in the shape transitional region 
of $^{142-152}$Nd and $^{144-154}$Sm. The calculations were carried out by using the SLy4 EDF parameterization
with a volume pairing strength  $V_{\rm n}=V_{\rm p}=-190\,{\rm MeV\,fm}^3$ in the model space of 
$N_{\rm sh}=20$ oscillator shells.
The pairing strength was adjusted to reproduce the experimental proton pairing gap of 1.23\,MeV in $^{142}$Nd.
In this realistic calculation we use $N_{\rm GH}=N_{\rm GL}=40$ and $N_{\rm Leg}=80$, which are the
recommended values based on recent analysis \cite{Stoitsov20131592}.
The FAM contour integration was carried out using a semicircle with  $R_{A_1}=200$\,MeV, discretized with $N_{A1}=12$ points.

Table~\ref{table:NdSm} summarizes the results.
The calculated ground state properties show a gradual  spherical-to-deformed shape transition with increasing neutron number. Moreover, in some of the isotopes we  predict pairing collapse. For that reason, the chosen set of nuclei is representative of a realistic situation encountered in global surveys across the nuclear landscape, where deformations and pairing may vary rapidly as a function of proton and neutron number.

The energy-weighted sum rules computed with the FAM agree well with
the HFB expressions of Appendix~\ref{sec:appendix}. This agreement holds regardless of nuclear  shape or pairing. As expected,
the energy-weighted sum rule for the isoscalar monopole operator increases with $N$
in the region of the shape transition; this is attributed to an increase of the root mean square radius with deformation. 
Similarly, the isoscalar quadrupole operator increases even more rapidly with increasing quadrupole deformation.

Next, we consider the energy weighted sum rules in constrained HFB states.
The  constrained HFB potential energy curve  as a function of quadrupole moment was obtained using  the quadratic  constraint.
The contribution from the  quadratic constraining potential was included consistently to the residual 
field in the FAM.
This kind of  calculation represents  local QRPA on top of constrained HFB   \cite{PhysRevC.82.064313}; it contains dynamical information about  non-equilibrium configurations in the deformation space.

\begin{figure}[htb]
\includegraphics[width=\columnwidth]{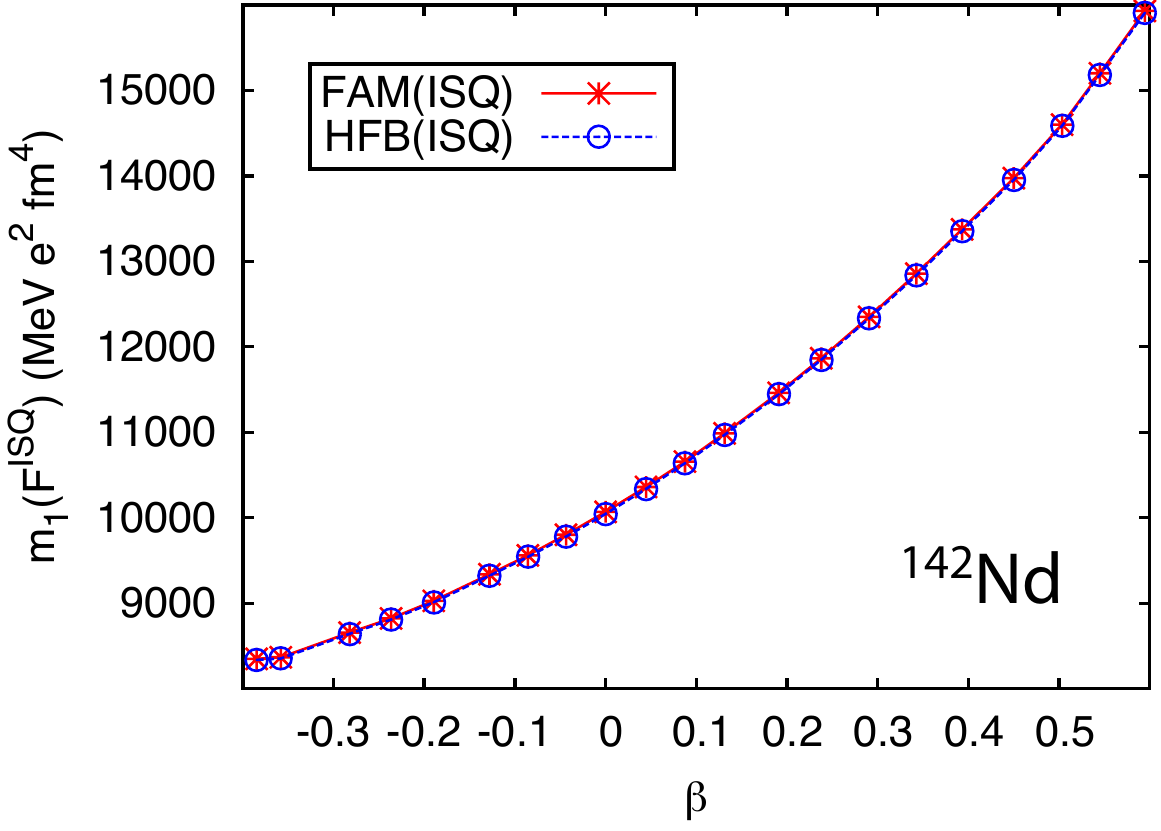}
  \caption{(Color online) Energy-weighted sum rule of the isoscalar quadrupole operator in $^{142}$Nd obtained 
  in the HFB  (solid  line with asterisks) and FAM (dashed line  with open circles) frameworks
  as a function of quadrupole deformation $\beta$ for the constrained HFB solutions. 
\label{fig:142Nd}}
\end{figure}

The energy-weighted sum rule of the isoscalar quadrupole operator as a function of quadrupole 
deformation is shown in Fig.~\ref{fig:142Nd}. The  sum rule 
increases monotonically with $\beta$  and
agrees very well with  HFB values.
This, together with results presented in Table~\ref{table:NdSm}, indicates that the Thouless theorem  provides a good 
approximation to the energy weighted sum rule within
the Skyrme-EDF picture, which is not based on the underlying Hamiltonian.

In passing, we should note that when departing from the HFB  minimum, there is a possibility of imaginary energy QRPA solutions; in such cases, a pair of QRPA poles would appear  on the imaginary axis. 
Although one expects no contribution to  odd-$k$ sum rules from such a pair, a careful consideration needs to be given to the   choice of integration  contour in the FAM. A general extension of the complex FAM formalism to the  case of local QRPA will be an interesting avenue for future studies.


\section{Conclusions \label{sec:conclusion}}

We propose an efficient formalism to compute sum rules by using the contour integration formalism within the complex-energy finite-amplitude method.
In particular when the order of the moment is odd, the obtained expression becomes extremely simple, as the sum rules 
appear as expansion coefficients of the Laurent series of the response function. 
The new formalism has been successfully benchmarked against the  matrix diagonalization method of QRPA.

We compare the energy-weighted sum rule obtained in the FAM with those based on
the Thouless theorem. Although the double commutator cannot be evaluated 
for general energy density functionals that are not based on a Hamiltonian,  the numerical results indicate that the 
theorem provides a very good approximation to $m_1$ when a large model space is employed and local gauge symmetry of the 
EDF is satisfied.
The inverse-energy-weighted sum rule was compared with the  constrained HFB result using the dielectric theorem, and  a perfect agreement 
was obtained regardless of the model space. 

Our results suggest that sum rules can be computed efficiently in the FAM even in cases when  other methods are not easily available (e.g., the Thouless theorem cannot be applied or constrained calculations cannot be carried out because of self-consistent symmetries assumed).
The extension of the  formalism to non-Hermitian operators is also straightforward,
as it has already been applied to the beta-decay rates with pn-FAM \cite{PhysRevC.90.024308}. 

The FAM approach to sum rules promises to add new functionality to the  EDF optimization framework of Refs.~\cite{PhysRevC.82.024313,PhysRevC.85.024304,PhysRevC.89.054314} as it will allow adding new kinds of data on multipole- and charge-exchange strength to the set of fit-observables defining the objective function.
The new FAM technique can be very useful when studying the nuclear response to non-trivial operators such as the nuclear Schiff moment,
which is closely related to the isoscalar dipole operator \cite{PhysRevC.72.045503,PhysRevC.89.014335}.

\begin{acknowledgments}

Useful discussions with J. Dobaczewski and T. Nakatsukasa
are gratefully acknowledged. 
This material is
based upon work supported by the U.S. Department of Energy, Office of
Science, Office of Nuclear Physics under Award Numbers No.\
DE-FG02-96ER40963 (University of Tennessee) and No.\ DE-SC0008511
(NUCLEI SciDAC Collaboration); by the NNSA's Stewardship Science
Academic Alliances Program under Award No.\ DE-NA0001820; by the Academy of Finland under the Centre of
Excellence Programme 2012--2017 (Nuclear and Accelerator
Based Physics Programme at JYFL); and the FIDIPRO
programme. An award of computer time was provided by the
Innovative and Novel Computational Impact on Theory and
Experiment (INCITE) program.
A part of the calculation was performed
in the resources of High Performance Computing Center, Institute for Cyber-Enabled Research, Michigan State University.
\end{acknowledgments}

\appendix
\section{Thouless theorem for monopole and quadrupole operators \label{sec:appendix}}

According to the Thouless theorem (\ref{eq:EWSR}), the energy-weighted sum rule for  isoscalar monopole and quadrupole operators of an  axially-deformed nucleus  are: 
  \begin{align}
  m_1({\rm ISM}) =& 4e^2\left(\frac{Z}{A}\right)^2 \frac{\hbar^2}{2m} A \langle r^2\rangle \, , \label{eq:m1-ism}\\
  m_1({\rm ISQ}) =&  e^2\left(\frac{Z}{A}\right)^2 \frac{\hbar^2}{2m} \frac{5}{2\pi}  A \langle r^2\rangle \left(1 + \sqrt{\frac{5}{4\pi}}\beta\right) \,. \label{eq:m1-isq}
  \end{align}
where $\langle r^2\rangle$ is the total  rms squared radius and $\beta$ is
the mass quadrupole deformation parameter:
\begin{align}
  \beta=&\sqrt{\frac{\pi}{5}}\frac{1}{A\langle r^2 \rangle}
  \int (3z^2-r^2)\rho(\bm{r})d\bm{r}.
\end{align}
For isovector operators, there appears an enhancement factor
\begin{align}
  \kappa = \frac{8m}{\hbar^2}(C^\tau_0 - C^\tau_1) \frac{\displaystyle\int |\nabla f(\bm{r})|^2 \rho_{\rm n}(\bm{r})\rho_{\rm p}(\bm{r}) d\bm{r}}{\displaystyle\int |\nabla f(\bm{r})|^2 \rho(\bm{r}) d\bm{r}},
\end{align}
where -- in the notation of Ref. \cite{PhysRevC.52.1827} -- $C^{\tau}_t$ is the coupling constant of the term $\rho_t\tau_t$ in the EDF.
The expressions for the isovector monopole and quadrupole operators are:
\begin{align}
  m_1({\rm IVM}) =& 4 e^2 \frac{\hbar^2}{2m} \frac{NZ}{A^2} \left[
    Z \langle r^2\rangle_{\rm n} + N\langle r^2\rangle_{\rm p}
    \right] (1+\kappa_{\rm IVM}), \\
  \kappa_{\rm IVM} =& \frac{8m}{\hbar^2} (C^\tau_0 - C^\tau_1)
  \frac{1}{A\langle r^2\rangle}
  \int r^2 \rho_{\rm n}(\bm{r}) \rho_{\rm p}(\bm{r})d\bm{r},
  \end{align}
and

  \begin{align}
  m_1({\rm IVQ}) =& e^2 \frac{\hbar^2}{2m}\frac{NZ}{A^2} \frac{5}{2\pi}
  \left[
    Z\langle r^2\rangle_{\rm n}\left( 1 + \sqrt{\frac{5}{4\pi}}\beta_{\rm n}\right) \right.   \nonumber \\
    + & \left.
    N\langle r^2\rangle_{\rm p}\left( 1 + \sqrt{\frac{5}{4\pi}}\beta_{\rm p}\right)
  \right] (1+\kappa_{\rm IVQ}),
  \\
  \kappa_{\rm IVQ} =& \frac{8m}{\hbar^2} (C^\tau_0 - C^\tau_1) \frac{1}{\displaystyle2A\langle r^2 \rangle
    \left( 1+\sqrt{\frac{5}{4\pi}}\beta \right)} \nonumber \\
 \times & \int (3z^2 + r^2) \rho_{\rm n}(\bm{r}) \rho_{\rm p}(\bm{r}) d\bm{r},
  \end{align}
where subscripts n/p indicate neutron/proton expectation values respectively.

\bibliographystyle{apsrev4-1}
\bibliography{sumrule}

\end{document}